\documentclass[aps,prb,twocolumn]{revtex4} 

\usepackage{graphicx}
\usepackage{dcolumn}
\usepackage{bm}

\newcommand{\comment}[1]{}

\begin{document}
\renewcommand{\theequation}{\arabic{section}.\arabic{equation}}

\title{
Quantum Ornstein-Zernike Equation}


\author{Phil Attard}
\affiliation{{\tt phil.attard1@gmail.com}}


\begin{abstract}
The non-commutativity of the position and momentum operators
is formulated as an effective potential in classical phase space
and expanded as a series of successive many-body terms,
with the pair term being dominant.
A non-linear partial differential equation in temperature and space
is given for this.
The linear solution is obtained explicitly,
which is valid at high and intermediate temperatures,
or at low densities.
An algorithm for solving the full non-linear problem is given.
Symmetrization effects accounting for particle statistics
are also written as a series of effective many-body potentials,
of which the pair term is dominant at terrestrial densities.
Casting these quantum  functions as pair-wise additive,
temperature-dependent, effective potentials
enables the established techniques of classical statistical mechanics
to be applied to quantum systems.
The quantum Ornstein-Zernike equation is given as an example.
\end{abstract}

\pacs{}

\maketitle

%
\section{Introduction}
\setcounter{equation}{0} \setcounter{subsubsection}{0}
%

Wigner\cite{Wigner32}
gave a formulation of quantum statistical mechanics
that expressed the probability density for classical phase space
as a multi-dimensional convolution integral
of the Maxwell-Boltzmann operator
acting on the momentum eigenfunctions.
He used this expression to obtain the primary quantum correction
to the classical free energy,
which was of second order in Planck's constant
for an unsymmetrized wave function.
Kirkwood\cite{Kirkwood33} showed that the temperature derivative
of Wigner's integrand was slightly more convenient
for developing expansions in powers of either Plank's constant
or of inverse temperature.
He used it to give both the first and second order quantum corrections
to the classical result for a symmetrized wave function.

I also have  developed a formulation
of  quantum statistical mechanics in classical phase space.
\cite{STD2,Attard18a}
This eschews Wigner's convolution integral,
but it nevertheless involves the same Maxwell-Boltzmann operator
acting on the momentum eigenfunctions.
I have named this phase function the commutation function,
because it accounts for the non-commutativity
of the position and momentum operators
that is otherwise neglected in classical mechanics.
Using Kirkwood's temperature derivative method,
I have obtained the quantum corrections up to
and including fourth order in Plank's constant.\cite{STD2,Attard18b}
The coefficients involve gradients of the potential energy,
and their number and complexity
grows exponentially for successive terms in the expansion.
Whilst undoubtedly useful at high temperatures,
where all systems are predominantly classical,
at intermediate and low temperatures the series expansion
is tedious to derive, complex to program, and slow to converge,
and it does not appear to be a practical approach
for condensed matter systems.

An alternative approach
is based upon a formally exact transformation
that expresses the commutation function
as a sum over energy eigenvalues and eigenfunctions.\cite{Attard18c}
Although these are not known explicitly for a general system,
they are known exactly for the simple harmonic oscillator.
Accordingly, in a mean field approach the commutation function
for a point in classical phase space
can be approximated by that of a simple harmonic oscillator
based on  a second order expansion of the potential
about the nearest local minimum for the configuration.
A Metropolis Monte Carlo simulation
using this mean field--simple harmonic oscillator approximation
has been shown to yield accurate results
at low temperatures for two condensed matter systems
where the quantum result is known to high accuracy:
a Lennard-Jones fluid,\cite{Hernando13,Attard18c}
and an harmonic crystal.\cite{Attard19}

A big advantage of the classical phase space formulation
of quantum statistical mechanics in general
is that the required computer time for a given statistical accuracy
scales sub-linearly with system size.
However the difficulty
with the mean field--simple harmonic oscillator approximation
is that one cannot be certain of its accuracy
in the absence of benchmark results for the same specific system.
Although the mean field aspect has been improved by applying
it at the pair level,\cite{Attard19}
it is unclear how to go beyond approximating the commutation function
as that of an effective simple harmonic oscillator.
Whilst undoubtedly there are strong arguments
in favor of mean field approaches in condensed matter,
there is some motivation to develop
a more systematic approach to evaluating the commutation function,
but without the tedium of the crude expansions
as derived by Wigner,\cite{Wigner32} by Kirkwood,\cite{Kirkwood33}
and by me.\cite{Attard18b}

This paper makes a significant advance
with a new, systematic expansion for the commutation function.
The explicit contributions are fewer and simpler
than in the high temperature or Planck's constant expansions,
while the approach avoids the \emph{ad hoc} nature
of the mean field--simple harmonic oscillator approximation.

Here the commutation function is written as an effective potential,
which is then expanded
as a formal series of singlet, pair, three-body functions  etc.
Postulating that the many body terms are often negligible,
only the pair commutation function needs to be retained
for the common case of a homogeneous system.
This is a function of the relative separations and the relative momenta
summed over all particle pairs.
A non-linear partial differential equation
is given for its temperature derivative
and spatial gradients.
An explicit solution is obtained in Fourier space
for the linearized equations,
which are valid at high and intermediate temperatures,
and also at large separations.
This provides a convenient starting point for the full non-linear solution
via step-wise integration down to low temperatures, if required.
As part of a computational algorithm,
the full solution for the pair function
can be stored on a three-dimensional grid
prior to the commencement of, say, a Monte Carlo computer simulation,
where it can be conveniently evaluated
by interpolation from the storage grid.
Beyond computer simulations,
casting the quantum commutation function
as a temperature-dependent, effective potential
lends itself to the standard analytic techniques
of classical statistical mechanics
such as density functional theory,
diagrammatic expansion,
integral equation methods,
asymptotic analysis,
etc.
As an example, the quantum Ornstein-Zernike equation is given.

%
\section{Analysis} \label{Sec:w2}
\setcounter{equation}{0} \setcounter{subsubsection}{0}
%

\subsection{Phase Space Formulation of Quantum Statistical Mechanics}

The classical phase space quantum probability density
for the canonical equilibrium system is
(see Refs~[\onlinecite{STD2,Attard18a}] and also Appendix \ref{Sec:Xi})
\begin{equation}
\wp({\bf p},{\bf q})
=
\frac{ e^{-\beta {\cal H}({\bf p},{\bf q})}}{N! h^{3N} Z(T)}
\, \omega({\bf p},{\bf q}) \, \eta^\pm({\bf p},{\bf q}).
\end{equation}
Here $N$ is the number of particles,
${\bf p} = \{{\bf p}_1,{\bf p}_2,\ldots,{\bf p}_N\}$
are their momenta,
with ${\bf p}_j = \{p_{jx},p_{jy},p_{jz}\}$
being the momentum of the $j$th particle
(three dimensional space is assumed),
 ${\bf q} = \{{\bf q}_1,{\bf q}_2,\ldots,{\bf q}_N\}$
are the positions,
$\beta = 1/k_\mathrm{B}T$ is the inverse temperature,
with $T$ the temperature
and $k_\mathrm{B}$ Boltzmann's constant,
and $h$ is Planck's constant.
The classical Hamiltonian is
${\cal H}({\bf p},{\bf q}) = {\cal K}({\bf p}) + U({\bf q})$,
with ${\cal K} = p^2/2m$ being the kinetic energy,
and $U$ being the potential energy.
The  partition function $Z(T)$ normalizes the phase space probability density.
The distinctly quantum aspects of the system
are embodied in the symmetrization function $\eta$
and the commutation function $\omega$.
Although it is straightforward to treat particle spin,\cite{Attard19}
in this paper it is not included.

The symmetrization and commutation functions
are defined in terms
of the unsymmetrized position and momentum eigenfunctions,
which in the position representation are respectively\cite{Messiah61}
\begin{equation}
|{\bf q}\rangle = \delta({\bf r}-{\bf q})
, \mbox{ and }
|{\bf p}\rangle
=
\frac{e^{-{\bf p}\cdot{\bf r}/i\hbar}}{V^{N/2} } .
\end{equation}
Here $\hbar=h/2\pi$,
and $V$ is the volume of the system.

The symmetrization function is formally\cite{Attard18a,Attard19}
\begin{equation}
\eta^\pm({\bf p},{\bf q})
\equiv
\frac{1}{\langle {\bf p} | {\bf q} \rangle }
\sum_{\hat{\mathrm P}} (\pm 1)^p \,
\langle \hat{\mathrm P} {\bf p} | {\bf q} \rangle ,
\end{equation}
with $\hat{\mathrm P}$  the permutation operator
and $p$ its parity.
The plus sign is for bosons and the minus sign is for fermions.

A pair-wise additive form of the symmetrization function
is given in \S \ref{Sec:Symm}.
See earlier work
for the derivation of the phase space probability density,
\cite{STD2,Attard18a}
and for the proof that the symmetrization factor
correctly accounts for boson and fermion statistics.\cite{Attard19,FN1}

\subsection{Commutation Function}

The commutation function $\omega$,
which is essentially the same as the functions
introduced by Wigner\cite{Wigner32}
and analyzed by Kirkwood,\cite{Kirkwood33}
is defined by
\begin{equation}\label{Eq:def-Wp}
e^{-\beta{\cal H}({\bf p},{\bf q})}
\omega({\bf p},{\bf q})
=
\frac{\langle{\bf q}|e^{-\beta\hat{\cal H}} |{\bf p}\rangle
}{\langle{\bf q}| {\bf p}\rangle } .
\end{equation}

The departure from unity of the commutation function
reflects the non-commutativity of the position and momentum operators.
To see this  simply note that \emph{if} position and momentum commuted,
\emph{then} the Maxwell-Boltzmann operator would factorize,
$e^{-\beta\hat{\cal H}({\bf r})}  =
 e^{-\beta U({\bf r})} e^{-\beta\hat{\cal K}({\bf r})}$.
Then since $e^{-\beta\hat{\cal K}} |{\bf p}\rangle
=e^{-\beta{\cal K}({\bf p})}  |{\bf p}\rangle$,
the right hand side would reduce to
the classical Maxwell-Boltzmann factor $e^{-\beta{\cal H}({\bf p},{\bf q})}$,
in which case $\omega({\bf p},{\bf q})=1$.

Obviously the system must become classical in the high temperature limit,
and so $\omega({\bf p},{\bf q}) \rightarrow 1$, $\beta \rightarrow 0$.

Further, the potential energy decays
to zero at large separations,
and the gradients of the potential are negligible
compared to the potential itself.
This means  that at large separations
the potential energy and the kinetic energy operator effectively commute,
and so one similarly must have
$\omega({\bf p},{\bf q}) \rightarrow 1$, all $ q_{jk} \rightarrow \infty$.

Following 
Kirkwood,\cite{Kirkwood33}
differentiation of the defining equation
with respect to inverse temperature gives\cite{STD2}
\begin{eqnarray} \label{Eq:dOmega/dbeta}
\frac{\partial \omega}{\partial \beta}
& = &
\frac{- \beta\hbar^2}{2m}  (\nabla^2 U) \omega
- \frac{\beta\hbar^2}{m} (\nabla U) \cdot (\nabla \omega)
\nonumber \\ & & \mbox{ }
 + \frac{\beta^2 \hbar^2}{2m} (\nabla U) \cdot (\nabla U) \omega
+ \frac{\hbar^2}{2m} \nabla^2 \omega
\nonumber \\ & & \mbox{ }
+ \frac{i\hbar}{m} {\bf p} \cdot  (\nabla \omega)
- \frac{i\hbar\beta}{m}  {\bf p} \cdot (\nabla U) \omega .
\end{eqnarray}
One can see from this that since $\nabla U \rightarrow 0$
at large separations,
the putative asymptote, $\omega({\bf p},{\bf q}) \rightarrow 1$,
makes the right hand side vanish,
which is consistent with the asymptote being a temperature-independent
constant.
This partial differential equation
provides the basis for the high temperature expansions
that were mentioned in the introduction.
\cite{Wigner32,Kirkwood33,STD2,Attard18b}
The multiplicity of gradients on the right hand side
is the origin for the rapid increase of number and complexity
of the contributions to each order of the expansions,
which are listed in Appendix \ref{Sec:Ww}.

Elsewhere
I have argued that it is more useful to cast
the commutation function as
an effective potential,
\cite{STD2}
\begin{equation}
W({\bf p},{\bf q}) \equiv \ln \omega({\bf p},{\bf q})
,\mbox{ or }
\omega({\bf p},{\bf q}) \equiv e^{W({\bf p},{\bf q})} .
\end{equation}
(The present notation differs from previous articles.)
In this form $W$ is a temperature-dependent effective potential.
The original rationale for this was
that $W$ is extensive with system size,
which is an exceedingly useful property in thermodynamics.
It was also hoped that the expansion of $W$
in powers of Planck's constant or inverse temperature
might have better convergence properties than those of $\omega$ itself,
a hope more happy in prospect than in actual experience.

\subsection{Many Body Expansion}

The effective potential $W$
provides the starting point for the new approach
to the commutation function,
which is the main point of this paper.

Defining $\tilde W \equiv W -\beta U$,
the temperature derivative
of the definition leads to\cite{STD2}
\begin{equation} \label{Eq:dW/dbeta}
\frac{\partial \tilde W}{\partial \beta }
=
- {\cal H}     
+
\frac{i\hbar}{m} {\bf p} \cdot \nabla \tilde W
 + \frac{\hbar^2}{2m}\nabla^2 \tilde W
+
\frac{\hbar^2}{2m}
\nabla \tilde W \cdot \nabla \tilde W ,
\end{equation}
with $\tilde W \sim -\beta {\cal H} $, $\beta \rightarrow 0$.
The right hand side contains
two terms linear in $\tilde W$
and one non-linear, quadratic term.

In general 
the potential energy is the sum of
one-body, two-body, three-body  etc.\ potentials,
\begin{eqnarray}
U({\bf q})
& = &
\sum_{j=1}^N u^{(1)}({\bf q}_{j})
+  \sum_{j<k}^N u^{(2)}({\bf q}_{j},{\bf q}_{k})
\nonumber \\ && \mbox{ }
+  \sum_{j<k<\ell}^N u^{(3)}({\bf q}_{j},{\bf q}_{k},{\bf q}_{\ell})
+\ldots
\end{eqnarray}
In this work three-body and higher potentials will be neglected.
It will also be assumed that there is no one-body potential,
that the system is homogeneous,
and that the pair potential is a function only of separation,
\begin{equation}
U({\bf q})
=
\sum_{j<k}^N u(q_{jk})
=
\frac{1}{2} \sum_{j,k}\!^{(k \ne j)} u(q_{jk}) ,
\end{equation}
where the particle separation 
is $q_{jk} = | {\bf q}_j - {\bf q}_k|$.

For the effective potential form of the commutation function,
a similar many body decomposition can be formally made,
with momentum now being included
\begin{eqnarray}
\lefteqn{
W({\bf p},{\bf q})
}  \\
& = &
\sum_{j=1}^N w^{(1)}({\bf p}_{j},{\bf q}_{j})
+  \sum_{j<k}^N w^{(2)}({\bf p}_{j},{\bf q}_{j};{\bf p}_k,{\bf q}_{k})
\nonumber \\ && \mbox{ }
+  \sum_{j<k<\ell}^N
w^{(3)}({\bf p}_{j},{\bf q}_{j};{\bf p}_{k},{\bf q}_{k};
{\bf p}_\ell,{\bf q}_{\ell})
+\ldots \nonumber
\end{eqnarray}
For the homogeneous case,
the singlet contribution vanishes,
($W^{(1)} = 0$,
$\tilde W^{(1)} = - \beta {\cal K}({\bf p})$,
and $\nabla \tilde W^{(1)}  = {\bf 0}$),
and the three-body and higher contributions,
will be neglected.
Assuming homogeneity, 
the pair commutation function for particles $j$ and $k$
is a function of the magnitudes of the relative momentum
and separation, and the angle between them,
$w^{(2)}(p_{jk},q_{jk},\theta_{jk})$.
Alternatively, 
the relative momentum can be aligned with the $z$ axis,
${\bf p}_{jk} = p_{jk}\hat{\bf z}$, and the separation
can be rotated to the $xz$-plane,
$w^{(2)}(p_{jk},q_{jk,x},q_{jk,z})$,
or
$w^{(2)}(p,q_\perp,q_\parallel)$.
In any case the commutation function that will be considered here is
\begin{equation}
W({\bf p},{\bf q})
=
\sum_{j<k}^N w^{(2)}({\bf p}_{jk},{\bf q}_{jk})
=
\frac{1}{2} \sum_{j,k}\!^{(k \ne j)} w^{(2)}_{jk}.
\end{equation}

Inserting this pair ansatz into the temperature derivative equation,
one sees that the left hand side,
and the constant and linear terms on the right hand side,
are all the sum of pair functions.
However the non-linear quadratic term on the right hand side
is
\begin{eqnarray} \label{Eq:DW.DW}
\lefteqn{
\nabla \tilde W \cdot \nabla \tilde W
} \nonumber \\
& = &
\sum_\ell \nabla_\ell \tilde W \cdot \nabla_\ell \tilde W
\nonumber \\ & = &
 \frac{1}{4} \sum_\ell
 \sum_{j,k}\!^{(k \ne j)} \sum_{j',k'}\!^{(k' \ne j')}
 \nabla_\ell  \tilde w^{(2)}_{jk} \cdot  \nabla_\ell  \tilde w^{(2)}_{j'k'}
\nonumber \\ & = &
 \frac{1}{4}
 \sum_{\ell,k,k'}\!^{(k,k' \ne \ell)}
 \nabla_\ell \tilde w^{(2)}_{\ell k}
 \cdot \nabla_\ell \tilde w^{(2)}_{\ell k'}
 \nonumber \\ && \mbox{ }
 +
 \frac{1}{4}
 \sum_{j,\ell,k'}\!^{(j,k' \ne \ell)}
 \nabla_\ell \tilde w^{(2)}_{j \ell} \cdot
 \nabla_\ell \tilde w^{(2)}_{\ell k'}
 \nonumber \\ && \mbox{ }
 +
 \frac{1}{4}
 \sum_{\ell,k,j'}\!^{(k,j' \ne \ell)}
 \nabla_\ell \tilde w^{(2)}_{\ell k} \cdot
 \nabla_\ell \tilde w^{(2)}_{j' \ell}
 \nonumber \\ && \mbox{ }
 +
 \frac{1}{4}
 \sum_{\ell,j,j'}\!^{(j,j' \ne \ell)}
 \nabla_\ell \tilde w^{(2)}_{j \ell} \cdot
 \nabla_\ell \tilde w^{(2)}_{j' \ell}
 \nonumber \\ & = &
 \sum_{j,k,k'}\!^{(k,k' \ne j)}
 \nabla_j \tilde w^{(2)}_{j k} \cdot
 \nabla_j \tilde w^{(2)}_{j k'} .
\end{eqnarray}
This is in essence a three-body term.
However, to the extent that the `force' on particle $j$ due to particle $k$
is uncorrelated with that due to $k'$, $k'\ne k$,
the sum of forces over $k'$ may be argued to be small,
perhaps averaging to zero.
In this case the $k'=k$ terms dominate,
and this becomes
\begin{eqnarray} \label{Eq:DW.DW2}
\nabla \tilde W \cdot \nabla \tilde W
& \approx &
 \sum_{j,k}\!^{(k \ne j)}
 \nabla_j \tilde w^{(2)}_{j k} \cdot
 \nabla_j \tilde w^{(2)}_{j k}
\nonumber \\ & = &
2 \sum_{j<k}
 \nabla_j \tilde w^{(2)}_{j k} \cdot
 \nabla_j \tilde w^{(2)}_{j k} .
\end{eqnarray}
This is now a two-body term,
which is expected to be accurate at low densities
and high temperatures where correlations are reduced.
The two-body solution
can test the magnitude of the neglected three-body terms,
and successively improve the pairwise additive approximation.

Applying the pair ansatz to the temperature derivative,
equating both sides term by term,
and dropping the superscript $(2)$, one obtains
\begin{eqnarray}
\frac{\partial \tilde w_{jk}}{\partial \beta }
& = &
- u_{jk} +
\frac{i\hbar}{m}
{\bf p}_{jk} \cdot \nabla_j \tilde w_{jk}
+ \frac{\hbar^2}{m} \nabla_j^2  \tilde w_{j k}
\nonumber \\ &&  \mbox{ }
+ \frac{\hbar^2}{m}
\nabla_j \tilde w_{j k} \cdot  \nabla_j \tilde w_{j k} .
\end{eqnarray}
Here and above
the symmetry $\tilde w_{j k} = \tilde w_{kj}$ has been exploited.
Note that $\tilde w_{j k} \sim -\beta u_{jk}$, $\beta \rightarrow 0$.

Write ${\bf p}_{jk} = p_z \hat{\bf z}$,
and  ${\bf q}_{jk} =  q_x \hat{\bf x} + q_z \hat{\bf z} $,
and $q_{jk} = \sqrt{q_x^2+q_z^2}$.
This then  becomes
\begin{eqnarray}
\frac{\partial \tilde w(p_z,q_x,q_z)}{\partial \beta }
& = &
- u(q)+
\frac{i\hbar}{m} p_z \tilde w_z
+ \frac{\hbar^2}{m} \left\{ \tilde w_{xx} + \tilde w_{zz} \right\}
\nonumber \\ &&  \mbox{ }
+ \frac{\hbar^2}{m} \left\{ \tilde w_x^2 + \tilde w_z^2 \right\}.
\end{eqnarray}
The subscripts on $\tilde w$ signify spatial derivatives.

Define the two dimensional Fourier transform pair
\begin{eqnarray}
\hat {\tilde w}(p_z,k_x,k_y)
& = &
\int \mathrm{d}{\bf q} \; e^{-i{\bf k} \cdot {\bf q}} \tilde w(p_z,q_x,q_z)
 \\ \nonumber
\tilde w(p_z,q_x,q_z)
& = &
\frac{1}{(2\pi)^2}
\int \mathrm{d}{\bf k} \; e^{i{\bf k} \cdot {\bf q}}
\hat {\tilde w}(p_z,k_x,k_y) .
\end{eqnarray}
The Fourier transform of a gradient such as $\tilde w_x({\bf q})$
is just $ik_x \hat{\tilde w}({\bf q})$.
Note that here  ${\bf q}$ is a two-dimensional vector.

The  Fourier transform of the temperature derivative is
\begin{eqnarray}
\lefteqn{
\frac{\partial \hat {\tilde w}(p_z,k_x,k_y)}{\partial \beta }
}  \\
& = &
- \hat u(k)
- \left[ \frac{\hbar}{m} p_z k_z
+ \frac{\hbar^2}{m} k^2 \right]
\hat{\tilde w}({\bf p},{\bf k})
\nonumber \\ & & \mbox{ }
- \frac{\hbar^2}{(2\pi)^2m}
\int \mathrm{d}{\bf k}'\;
{\bf k}' \cdot ({\bf k}-{\bf k}')\,
\hat{\tilde w}({\bf p},{\bf k}')\,
\hat{\tilde w}({\bf p},{\bf k}-{\bf k}').\nonumber
\end{eqnarray}
Here $k^2 = k_x^2 + k_z^2 $.
Write
$b({\bf k})\equiv -\hbar  p_z k_z /m -  \hbar^2 k^2/m $,
which is negative for large $k$.

\subsubsection{Linear Solution}

In the case that $w$ is small,
as occurs at high temperatures or at large separations,
one can  simply neglect the quadratic term,
so that the differential equation is linear,
${\partial \hat {\tilde w}_\mathrm{lin}}/{\partial \beta }
= - \hat u + b \hat {\tilde w}_\mathrm{lin}$.
This has  solution
\begin{equation}
\hat {\tilde w}_\mathrm{lin}^{(2)}({\bf k})
= \frac{- \hat u(k)}{b({\bf k})} [ e^{\beta b({\bf k})} - 1 ].
\end{equation}
The linear approximation is valid at high
and  intermediate temperatures,
or at large separations.
The pair ansatz is exact in the linear regime.

This  result
significantly improves upon the high temperature expansions,
with  a larger regime of validity
and being much simpler to evaluate and to analyze.

In the limit of large $k$,
this gives
$\hat {\tilde w}_\mathrm{lin} \sim {\hat u(k)}/{b(k)}
\sim  k^{-2} \hat u(k) $,
$ k \rightarrow \infty$.
This says that at small separations
$w^{(2)}(q)$  dominates the classical $-\beta u^{(2)}(q)$,
which is essential because the quantum effects of non-commutativity
have to dominate on small lengths scales.
(Eg.\ the electron-nucleus interaction,
where the classical Maxwell-Boltzmann weight alone
would lead to catastrophe.)

For small $k$, $b(k) \rightarrow 0$,
and $\hat {\tilde w}_\mathrm{lin}(k) \sim  -\beta \hat u(k)$.
This confirms that $w(q)$ decays more quickly at large separations
than the pair potential itself.
(This can be shown to be true in the non-linear case as well.)

The linear solution can be inserted into the convolution
integral and the temperature integration performed explicitly.
This  gives the first non-linear correction
and tells how reliable the linear solution is.

For any value of $p_{jk}=p_{jk,z} $,
using the two-dimensional fast Fourier transform,
one can  numerically invert
$\hat{\tilde w}_\mathrm{lin}^{(2)}(p_{jk};k_x,k_z)$
giving $\tilde w_\mathrm{lin}^{(2)}(p_{jk,z},q_{jk,x},q_{jk,z})$.
Summing over pairs gives the full commutation function
for any phase space point, $W_\mathrm{lin}({\bf p},{\bf q})$.
Obviously if many phase space points are required,
it would be most efficient to evaluate
$\tilde w_\mathrm{lin}^{(2)}(p,q_\perp,q_\parallel)$
once only,
storing this on a three-dimensional grid.
The full commutation function can be evaluated
for any phase space point by summing over all pairs
each interpolated value of the stored function.

\subsubsection{Non-Linear Solution}

The non-linear partial differential equation
can be solved by stepping forward in inverse temperature,
starting from a high temperature with the linear solution,
\begin{equation}
\hat {\tilde w}^{(2)}({\bf k};\beta_{\alpha+1}) =
\hat {\tilde w}^{(2)}({\bf k};\beta_\alpha)
+\Delta \beta
\frac{\partial \hat {\tilde w}^{(2)}({\bf k};\beta_\alpha)
}{\partial \beta } ,
\end{equation}
\comment{
The non-linear temperature derivative is used at each temperature node,
\begin{eqnarray}
\lefteqn{
\frac{\partial \hat {\tilde w}_\alpha^{(2)}({\bf p},{\bf k})}{\partial \beta }
}  \\
& = &
- \hat u({\bf k})
- \left[ \frac{\hbar}{m} p_z k_z
+ \frac{\hbar^2}{m} k^2 \right]
\hat{\tilde w}_\alpha^{(2)}({\bf k})
\nonumber \\ & & \mbox{ }
- \frac{\hbar^2}{(2\pi)^2m}
\int \mathrm{d}{\bf k}'\;
{\bf k}' \cdot ({\bf k}-{\bf k}')\,
\hat{\tilde w}_\alpha^{(2)}({\bf k}')\,
\hat{\tilde w}_\alpha^{(2)}({\bf k}-{\bf k}').\nonumber
\end{eqnarray}
} 
Runge-Kutta procedures can be used
to accelerate and to stabilize the temperature integration.
Either evaluate the convolution
integral with the two-dimensional fast Fourier transform,
or else use finite differences in real space.
As in the linear case,
precalculate the non-linear
$w^{(2)}(p,q_\perp,q_\parallel)$
and store it on a three-dimensional grid.

\subsubsection{Singlet Potential}

Including a singlet potential leads to
\begin{eqnarray} \label{Eq:tw1j}
\frac{\partial \tilde w^{(1)}_j}{\partial \beta}
& = &
\frac{-p_j^2}{2m}  
-u^{(1)}_j
+ \frac{i\hbar}{m} {\bf p}_j \cdot \nabla_j  \tilde w^{(1)}_j
+ \frac{\hbar^2}{2m} \nabla_j^2  \tilde w^{(1)}_j
 \nonumber \\ &  & \mbox{ }
+ \frac{\hbar^2}{2m} \nabla_j  \tilde w^{(1)}_j
\cdot \nabla_j  \tilde w^{(1)}_j,
\end{eqnarray}
and
\comment{ 
\begin{eqnarray}
\frac{\partial \tilde w_{jk}^{(2)}}{\partial \beta }
& = &
- u_{jk}^{(2)}
+
\frac{i\hbar}{m}
{\bf p}_{jk} \cdot \nabla_j \tilde w^{(2)}_{jk}
+ \frac{\hbar^2}{m} \nabla_j^2  \tilde w^{(2)}_{j k}
\nonumber \\ &  & \mbox{ }
+ \frac{\hbar^2}{m}
\nabla_j \tilde w^{(2)}_{j k} \cdot  \nabla_j \tilde w^{(2)}_{j k}
  \\ &  & \mbox{ }
+
\frac{\hbar^2}{m}
\left\{ \nabla_j \tilde w^{(1)}_j - \nabla_k \tilde w^{(1)}_k \right\}
\cdot \nabla_j  \tilde w^{(2)}_{jk} .
\nonumber
\end{eqnarray}
} 
\begin{eqnarray}  \label{Eq:tw2jk}
\frac{\partial \tilde w_{jk}^{(2)}}{\partial \beta }
& = &
- u_{jk}^{(2)}
+
\frac{i\hbar}{2m} \left\{
{\bf p}_{j} \cdot \nabla_j \tilde w^{(2)}_{jk}
+
{\bf p}_{k} \cdot \nabla_k \tilde w^{(2)}_{jk}
\right\}
\nonumber \\ && \mbox{ }
+ \frac{\hbar^2}{2m}
 \left\{ \nabla_j^2  \tilde w^{(2)}_{j k}
+ \nabla_k^2  \tilde w^{(2)}_{j k} \right\}
 \\ && \mbox{ }
+ \frac{\hbar^2}{2m}
 \left\{
\nabla_j \tilde w^{(2)}_{j k} \cdot  \nabla_j \tilde w^{(2)}_{j k}
+ \nabla_k \tilde w^{(2)}_{j k} \cdot  \nabla_k \tilde w^{(2)}_{j k}
\right\}
\nonumber \\ && \mbox{ }
+
\frac{\hbar^2}{m}
\left\{
\nabla_j \tilde w^{(1)}_j \cdot \nabla_j  \tilde w^{(2)}_{jk}
+
\nabla_k \tilde w^{(1)}_k \cdot \nabla_k  \tilde w^{(2)}_{jk}
\right\}, \nonumber
\end{eqnarray}
where
$w_{jk}^{(2)} = w^{(2)}({\bf p}_j,{\bf q}_j;{\bf p}_k,{\bf q}_k)
=w_{kj}^{(2)}$.

\subsection{Symmetrization Function} \label{Sec:Symm}

Decomposing the permutations into loops,
the symmetrization function is\cite{STD2,Attard18a,Attard19}
\begin{eqnarray}
\lefteqn{
\eta^\pm({\bf p},{\bf q})
}  \\
& \equiv &
\frac{1}{\langle {\bf p} | {\bf q} \rangle }
\sum_{\hat{\mathrm P}} (\pm 1)^p \,
\langle \hat{\mathrm P} {\bf p} | {\bf q} \rangle
\nonumber \\ & = &
\frac{1}{\langle {\bf p} | {\bf q} \rangle }
\left\{
\langle {\bf p} | {\bf q} \rangle
\pm \sum_{j,k} \!' \langle \hat{\mathrm P}_{jk}{\bf p} | {\bf q} \rangle
+ \sum_{j,k,\ell} \!'  \langle \hat{\mathrm P}_{jk}
\hat{\mathrm P}_{k\ell}{\bf p}
| {\bf q} \rangle
\right. \nonumber \\ && \mbox{ } \left.
+ \sum_{j,k,\ell,m} \!\!\!'\;
\langle \hat{\mathrm P}_{jk}\hat{\mathrm P}_{\ell m} {\bf p}
 | {\bf q} \rangle
\pm \ldots
\right\}
\nonumber \\ & = &
1 + \sum_{j<k} \eta^{\pm(2)}_{jk}
+ \sum_{j<k<\ell} \eta^{\pm(3)}_{jk\ell}
+ \sum_{j,k,\ell,m} \!\!\!'\;
\eta^{\pm(2)}_{jk}\eta^{\pm(2)}_{\ell m}
\pm \ldots \nonumber
\end{eqnarray}
The prime on the summations indicates that all the labels are different
\emph{and} that each  permutation is counted once only.
Here $\hat{\mathrm P}_{jk}$ is the transposition of particles $j$ and $k$,
which has odd parity.
The loops are successive transpositions of particles.
For example, a three-loop or trimer is
$  \hat{\mathrm P}_{jk}\hat{\mathrm P}_{k\ell}
\{ j,k,\ell\} = \{\ell, j,k\}$.
For $N$ objects,
the number of distinct permutations
consisting of $m_\ell$  $\ell$-loops
(ie.\ $\sum_{\ell=1}^N \ell m_\ell = N$) is
${N!}/{\prod_\ell  \ell^{m_\ell}m_\ell!}$.

With 
$\langle {\bf q}| {\bf p} \rangle
= V^{-N/2} \prod_{j=1}^N e^{-{\bf p}_j \cdot {\bf q}_j /i\hbar}$,
the dimer symmetrization loop
for two specific particles is
\begin{eqnarray}
\eta^{\pm(2)}_{jk}
& \equiv &
\frac{\pm\langle \hat{\mathrm P}_{jk}{\bf p} | {\bf q} \rangle
}{
\langle {\bf p} | {\bf q} \rangle }
\nonumber \\ & = &
\frac{\pm e^{{\bf p}_k \cdot {\bf q}_j /i\hbar}
e^{{\bf p}_j \cdot {\bf q}_k /i\hbar}
}{
e^{{\bf p}_j \cdot {\bf q}_j /i\hbar}
e^{{\bf p}_k \cdot {\bf q}_k /i\hbar} }
\nonumber \\ & = &
 \pm e^{[{\bf p}_k - {\bf p}_j ] \cdot {\bf q}_j /i\hbar}
 e^{[{\bf p}_j - {\bf p}_k ] \cdot {\bf q}_k /i\hbar}
\nonumber \\ & = &
\pm e^{-{\bf p}_{jk}  \cdot {\bf q}_{jk}/i\hbar} .
\end{eqnarray}
Similarly the trimer  loop for three specific particles is
\begin{eqnarray}
\eta^{\pm(3)}_{jk\ell}
& \equiv &
\frac{\langle \hat{\mathrm P}_{jk}\hat{\mathrm P}_{k\ell}{\bf p}
 | {\bf q} \rangle
}{
\langle {\bf p} | {\bf q} \rangle }
\nonumber \\ & = &
\frac{ e^{{\bf p}_\ell \cdot {\bf q}_j /i\hbar}
e^{{\bf p}_j \cdot {\bf q}_k /i\hbar}
e^{{\bf p}_k \cdot {\bf q}_\ell /i\hbar}
}{
e^{{\bf p}_j \cdot {\bf q}_j /i\hbar}
e^{{\bf p}_k \cdot {\bf q}_k /i\hbar}
e^{{\bf p}_\ell \cdot {\bf q}_\ell /i\hbar} }
\nonumber \\ & = &
 e^{{\bf p}_{\ell j}  \cdot {\bf q}_j /i\hbar}
 e^{{\bf p}_{j k}  \cdot {\bf q}_k /i\hbar}
 e^{{\bf p}_{k \ell}  \cdot {\bf q}_\ell /i\hbar}
\nonumber \\ & = &
 e^{{\bf q}_{\ell j}  \cdot {\bf p}_j /i\hbar}
 e^{{\bf q}_{j k}  \cdot {\bf p}_k /i\hbar}
 e^{{\bf q}_{k \ell}  \cdot {\bf p}_\ell /i\hbar} .
\end{eqnarray}
In general a specific $l$-loop is
\begin{equation}
\eta^{\pm(\ell)}_{j_1,j_2,\ldots ,j_\ell}
=
 e^{{\bf p}_{ j_{1} j_\ell}  \cdot {\bf q}_{j_\ell} /i\hbar}
\prod_{k=1}^{\ell-1}
 e^{{\bf p}_{j_{k+1} j_k }  \cdot {\bf q}_{j_k} /i\hbar} .
\end{equation}

The sums of these specific loops are phase functions,
$ \eta^{\pm(2)}({\bf p},{\bf q}) \equiv \sum_{j<k} \eta^{\pm(2)}_{jk}$,
$ \eta^{\pm(3)}({\bf p},{\bf q}) \equiv
\sum_{j<k<\ell} \eta^{\pm(3)}_{jk\ell}$,
etc.
(This notation differs from that of Ref.~[\onlinecite{STD2}].)

Because the complex exponentials  are highly oscillatory,
they average to zero unless the exponents are small.
Hence the loops are compact in phase space,
and the phase functions are extensive,
$  \eta^{\pm(\ell)}({\bf p},{\bf q}) = {\cal O}(N)$.

The fourth term in the expansion written out explicitly above
contains the products of two dimer loops.
They are not actually independent because all labels must be different.
Since each permutation appears once only,
it can be written as half the product of independent dimer functions
plus a lower order correction,
\begin{eqnarray}
\lefteqn{
\sum_{j,k,\ell,m} \!\!\!'\;
\eta^{\pm(2)}_{jk}\eta^{\pm(2)}_{\ell m}
} \nonumber \\
& = &
\frac{1}{8}
\sum_{j,k,\ell,m}
\overline \eta^{\pm(2)}_{jk}
\overline \eta^{\pm(2)}_{\ell m}
\overline \delta_{j\ell}
\overline \delta_{jm}
\overline \delta_{k\ell}
\overline \delta_{km}
\nonumber \\ & = &
\frac{1}{8}\!
\sum_{j,k,\ell,m}\!\!\!
\overline \eta^{\pm(2)}_{jk}
\overline \eta^{\pm(2)}_{\ell m}
-\frac{4}{8}
\sum_{j,k,\ell}\!
\overline \eta^{\pm(2)}_{jk}
\overline \eta^{\pm(2)}_{j\ell}
+\frac{2}{8}
\sum_{j,k}\!
(\overline \eta^{\pm(2)}_{jk} )^2
\nonumber \\ & = &
\frac{1}{2} \eta^{\pm(2)}({\bf p},{\bf q})^2 + {\cal O}(N),
\end{eqnarray}
where $\overline \delta_{jk} \equiv 1 - \delta_{jk} $
and $\overline \eta^{\pm(2)}_{jk} \equiv
\overline \delta_{jk}  \eta^{\pm(2)}_{jk} $.
Analogous results hold for all the products of loops
that appear in the loop permutation expansion.
Keeping only the leading order of independent products in each case,
the symmetrization function can be written
\begin{eqnarray}
\lefteqn{
\eta^\pm({\bf p},{\bf q})
} \nonumber \\
& \approx &
1 +  \eta^{\pm(2)}({\bf p},{\bf q})
+  \eta^{\pm(3)}({\bf p},{\bf q})
+ \frac{1}{2}  \eta^{\pm(2)}({\bf p},{\bf q})^2
+ \ldots
\nonumber \\ & = &
\exp
\left\{
\eta^{\pm(2)}({\bf p},{\bf q}) +  \eta^{\pm(3)}({\bf p},{\bf q})
 +  \eta^{\pm(4)}({\bf p},{\bf q})
+ \ldots \right\}
\nonumber \\ & = &
\prod_{\ell=2}^\infty e^{ \eta^{\pm(\ell)}({\bf p},{\bf q}) } .
\end{eqnarray}
The symmetrization loop functions
that are retained in the exponent are extensive, ${\cal O}(N)$.

This loop expansion and exponential form for the symmetrization function
has the same character as earlier work
in which the quantum grand potential
was expressed as a series of loop potentials.\cite{STD2,Attard18a}
The main differenc is that here
the exponentiation of the series of extensive single loops
is done for the phase function itself
rather than after statistical averaging.


\subsection{Generalized Mayer-$f$ Function}

The symmetrization loop phase functions
are related to the many body formulation of the commutation function.
Retaining only the pair term for the latter 
is valid at high temperatures or low densities.
Retaining only the dimer term in the symmetrization function
is valid at low densities.
The quantitative meaning of `high' and `low'
obviously depends on the particular system,
but it seems likely to encompass
most of condensed matter at terrestrial densities and temperatures.

Assuming a pairwise additive potential
(and no singlet potential),
and keeping only the pair terms in both the commutation
and symmetrization functions,
the grand canonical partition function is
(see Appendix \ref{Sec:Xi})
\begin{eqnarray}
\Xi
& \approx &
 \sum_{N=0}^\infty \frac{e^{\beta \mu N}}{N! h^{3N}}
 \int \mathrm{d}{\bf \Gamma}\;
e^{-\beta {\cal H}({\bf \Gamma})}
e^{W({\bf \Gamma})}
\prod_{\ell=2}^\infty e^{ \eta^{\pm(\ell)}({\bf \Gamma}) }
\nonumber \\ & \approx &
 \sum_{N=0}^\infty \frac{1}{N! h^{3N}}
 \int \mathrm{d}{\bf \Gamma}\;
 \prod_{j=1}^N \left[ e^{\beta \mu} e^{-\beta p_j^2/2m } \right]
 \nonumber \\ &  & \mbox{ } \times
 \prod_{j<k}^N
 \left[
 e^{-\beta u_{jk}^{(2)}}  e^{w_{jk}^{(2)}} e^{\eta^{\pm(2)}_{jk}}
 \right]
\nonumber \\ & \equiv &
 \sum_{N=0}^\infty \frac{1}{N! h^{3N}}
 \int \mathrm{d}{\bf \Gamma}\;
  \prod_{j=1}^N  z_j \;
 \prod_{j<k}^N  \left[ 1 + f_{jk}^{(2)}   \right].
\end{eqnarray}
Here ${\bf \Gamma} = \{ {\bf p},{\bf q}\}$
is a point in classical phase space,
and $z_j \equiv  e^{\beta \mu} e^{-\beta p_j^2/2m } $
is a generalized fugacity.
The quantity
$f_{jk}^{(2)} \equiv f^{(2)}({\bf p}_{jk},{\bf q}_{jk})
= e^{-\beta u_{jk}^{(2)}}  e^{w_{jk}^{(2)}} e^{\eta^{\pm(2)}_{jk}} - 1$
is a generalized pair Mayer-$f$ function
that depends upon the relative position and momentum
of the two particles.
It has the desirable property that
$ f^{(2)}({\bf p}_{jk},{\bf q}_{jk}) \rightarrow 0$,
$q_{jk} \rightarrow \infty$.
(If there is a singlet potential,
then its Maxwell-Boltzmann factor would be included in the fugacity,
along with the singlet part of the commutation function.
The pair part of the commutation function would need to be modified.)

This generalized Mayer-$f$ function
allows quantum systems to be treated with
the powerful techniques that have advanced
the field of classical statistical mechanics.
Examples include
cluster diagrams,
density functional theory,
integral equation methods,
and asymptotic analysis.
\cite{Pathria72,Hansen86,TDSM}
The  Mayer-$f$ function and
cluster diagrams are not restricted to pair-wise additive potentials,
\cite{Morita61,Stell64,Attard92}
although this is certainly the most common case.

For example,
the quantum pair Ornstein-Zernike equation
can just be  written down,
\begin{equation}
h({\bf \Gamma}_1,{\bf \Gamma}_2)
=
c({\bf \Gamma}_1,{\bf \Gamma}_2)
+
\int  \mathrm{d}{\bf \Gamma}_3\;
\rho({\bf \Gamma}_3)
c({\bf \Gamma}_1,{\bf \Gamma}_3)
h({\bf \Gamma}_3,{\bf \Gamma}_2) ,
\end{equation}
where ${\bf \Gamma}_j = \{ {\bf p}_j, {\bf q}_j \}$,
the singlet density is
$\rho^{(1)}_j \equiv \rho({\bf \Gamma}_j) =
N e^{-\beta p_j^2/2m } /(2\pi m/\beta)^{3/2}V$,
$h_{12}=g_{12}-1$ is the total correlation function,
$c_{12}$ is the direct correlation function,
and the pair distribution function is related to the pair density as
$\rho^{(2)}_{12} = \rho^{(1)}_1 \rho^{(1)}_2 g_{12}$.
This may be combined with, for example,
the hypernetted chain approximation,
\begin{equation}
h_{12}
=
-1 + 
e^{ h_{12} - c_{12}
-\beta u_{12}^{(2)} + w_{12}^{(2)} + \eta^{\pm(2)}_{12} },
\end{equation}
to give a closed system of equations.
The asymptote is
\begin{equation}
c_{12} \sim
-\beta u_{12}^{(2)} + w_{12}^{(2)} + \eta^{\pm(2)}_{12} ,
\;\; q_{12} \rightarrow \infty.
\end{equation}

One should be aware that the various quantities are complex.
(Recall that these are quantum probabilities,
albeit in classical phase space.)
Because of the oscillations at large separations,
it may be worth  adding numerically a damping or cut-off factor,
and checking that the final results are independent
of its precise value or form.
One should also be aware that
the requisite commutation function can vary with the quantity
being averaged.\cite{Attard18a}


%
\section{Conclusion}
\setcounter{equation}{0} \setcounter{subsubsection}{0}
%

In this paper a practical method has been given
for calculating the commutation function
that is suitable for terrestrial condensed matter systems.
This function gives the proper weight to phase space points
by accounting for the non-commutativity of position and momentum.

Previously most work has been focussed on obtaining
a few terms in an expansion in powers of Planck's constant
or inverse temperature.\cite{Wigner32,Kirkwood33}
The problem with this is that the coefficients
involve increasingly higher order gradients of the potential
and their products,
and they grow quickly in number and complexity
as more terms are retained in the series.\cite{STD2,Attard18a,Attard18b}
An alternative approach involves a mean field calculation
that invokes the exact simple harmonic oscillator commutation function,
albeit one appropriate for a second order expansion
of the potential energy for the instantaneous configuration.\cite{Attard18c}
The limitation of this approximation is that it is not obvious
how to systematically improve it.
A third possibility is to  formally write the commutation function
as a sum over energy eigenfunctions,
but this is not practically useful in the general case
where these are unknown.
Indeed, one great advantage of the classical phase space
formulation of quantum statistical mechanics
is that it avoids having to obtain the energy eigenvalues and eigenfunctions.

The approach developed in the present paper
is to cast the commutation function as a
temperature-dependent effective potential.
This is then written as a series of many-body terms,
which can be systematically truncated at any desired order
depending on the needs of a particular system or algorithm.

The simplest approximation is to retain only the pair term,
which is explored in detail here.
(The singlet term vanishes for a homogeneous system.)
In this case there is a non-linear partial differential equation
in temperature and position
for the pair commutation function.
At high and intermediate temperatures,
or at large separations,
this can be linearized and solved explicitly in Fourier space.
This can be used directly,
or else it provides a starting point for solving the non-linear equation
by Runge-Kutta methods, for example.
It appears feasible to use the method
in typical computer approaches to classical statistical mechanics,
such as Monte Carlo simulation methods.
Because the commutation function is pair-wise additive,
one can pre-calculate it and store it on a three-dimensional grid.
This enables the full commutation function to be evaluated
for any configuration by summing the interpolated  values
over all relevant pairs.

In addition to the commutation function,
the symmetrization function has here
also been cast as an effective potential
by invoking a loop expansion.
This has similarities to earlier work
which expressed the quantum grand potential
as a series of loop potentials.\cite{STD2,Attard18a}
Truncating the loop expansion at the dimer level
gives an effective pair potential that can be combined
with that for the commutation function and the actual potential energy
to create a pair Mayer-$f$ function.
With this most of the well-known results
of classical equilibrium statistical mechanics
can be directly applied to quantum systems.

Although the present results are formulated in classical phase space,
ultimately they are quantum in nature,
which requires considerations beyond
what is usual in classical statistical mechanics.
For example, the phase space weight
can vary with the quantity being averaged,\cite{Attard18a}
and it is complex,
as befits a quantum probability,
since the commutation and symmetrization functions
have imaginary part odd in momentum,
$W({\bf p},{\bf q})^* = W(-{\bf p},{\bf q})$, and
$\eta({\bf p},{\bf q})^* = \eta(-{\bf p},{\bf q})$.
The oscillatory nature of these poses challenges
in any numerical quadrature.
The incorporation of particle spin is a further
quantum feature not present in classical systems.\cite{Attard19}
Despite these and no doubt other practical challenges,
the present results
for the commutation and symmetrization functions
indicate the path
for quantum statistical mechanics in classical phase space.

{\bf Note Added.}
Numerical work after publication indicates that it is feasible
to calculate accurately the singlet commutation function as outlined here
for the case of the simple harmonic oscillator.
For the case of the pair commutation function for Lennard-Jones particles
further work is required to improve the reliability at lower temperatures.
Use in Metropolis Monte Carlo simulations has proven tractable.
\textbf{Post Scriptum.}
See Appendix~\ref{Sec:Results} for belated numerical results.



\appendix

%
\section{Grand Partition Function} \label{Sec:Xi}
\setcounter{equation}{0} \setcounter{subsubsection}{0}
\renewcommand{\theequation}{\Alph{section}.\arabic{equation}}
%

The expression for the quantum grand partition function
in classical phase space follows directly from
the von Neumann trace as a sum over quantum states,
the formal symmetrization of the wave function,
and the completeness of the position and momentum states.
It is\cite{Attard18a}
\begin{eqnarray}
\Xi^\pm
& = &
\mbox{TR}_{\in {\cal U}} \left\{ e^{-\beta \hat{\cal H}}  \right\}
\nonumber \\ & = &
\sum_{N=0}^\infty \frac{z^N}{N!}
\sum_{\hat{\mathrm P}} (\pm 1)^p
\sum_{{\bf p}\in {\cal N}}
\langle \hat{\mathrm P} {\bf p} | e^{-\beta \hat{\cal H}} |  {\bf p} \rangle
 \nonumber \\ & = &
\sum_{N=0}^\infty \frac{z^N}{N!}
\sum_{\hat{\mathrm P}} (\pm 1)^p
\sum_{{\bf p}\in {\cal N}}
 \int  \mathrm{d}{\bf q} \;
\langle \hat{\mathrm P} {\bf p} |  {\bf q} \rangle \,
\langle {\bf q} | e^{-\beta \hat{\cal H}} | {\bf p} \rangle
\nonumber \\ & = &
\sum_{N=0}^\infty \frac{z^N}{h^{dN} N!}
\sum_{\hat{\mathrm P}} (\pm 1)^p
\! \int \!\! \mathrm{d}{\bf \Gamma} \,
\frac{ \langle {\bf q} | e^{-\beta \hat{\cal H}} | {\bf p} \rangle
}{ \langle {\bf q} |  {\bf p} \rangle }
\frac{ \langle \hat{\mathrm P} {\bf p} |  {\bf q} \rangle
}{ \langle {\bf p} |  {\bf q} \rangle }
\nonumber \\ & \equiv &
\sum_{N=0}^\infty \frac{z^N}{h^{dN} N!}
\int \mathrm{d}{\bf \Gamma}\;
e^{-\beta{\cal H}({\bf \Gamma})}
\omega({\bf \Gamma}) \eta^\pm({\bf \Gamma}) .
\end{eqnarray}
Here $z = e^{\beta \mu}$ is the fugacity,
the dimensionality is usually $d=3$,
and ${\bf \Gamma} = \{{\bf p},{\bf q}\}$
is a point in classical phase space.

The first equality here is the
von Neumann trace form for the partition function.
\cite{Messiah61,Merzbacher70,Pathria72}
The sum is over allowed unique states:
each distinct state can only appear once.

The second equality writes the trace as a sum over \emph{all} momentum states,
symmetrizing the eigenfunctions.\cite{Attard18a}
This formulation of particle statistics is formally exact,
and carries state occupancy rules over to the continuum.

The third equality  inserts the completeness condition
$\int \mathrm{d}{\bf q}\; |{\bf q} \rangle \, \langle {\bf q} |
=\delta({\bf r}'- {\bf r}'')$,
to the left of the Maxwell-Boltzmann operator.
This produces an asymmetry in position and momentum
that is discussed elsewhere.\cite{Attard18a}

The fourth equality transforms to the momentum continuum.
The factor from the momentum volume element, $\Delta_p^{-dN}
= (2\pi\hbar/L)^{-dN}$,
combines with the factor of $V^{-N}= L^{-dN}
=\langle {\bf q} |  {\bf p} \rangle \, \langle {\bf p} |  {\bf q} \rangle $
to give the prefactor $h^{-dN}$.
This is now an integral over classical phase space.

The fifth equality writes the phase space integral
in terms of  the commutation function $\omega$,
and symmetrization function $\eta^\pm$,
as were used in the text.

%
\section{Expansions for the Commutation Function} \label{Sec:Ww}
\setcounter{equation}{0} \setcounter{subsubsection}{0}
\renewcommand{\theequation}{\Alph{section}.\arabic{equation}}
%

\subsection{Series Expansion for $\omega$}

From the temperature derivative,
Eq.~(\ref{Eq:dOmega/dbeta}),
Kirkwood\cite{Kirkwood33} derived a recursion relation for the coefficients
in an expansion of the commutation function $\omega$
in powers of Planck's constant.
A similar procedure was followed by me
with an expansion in powers of inverse temperature,\cite{Attard18b}
\begin{equation}
\omega
=
\sum_{n=0}^\infty \omega_n \beta^n .
\end{equation}
One has $\omega_0=1$, which is the classical limit,
and $\omega_1=0$, since there are no terms of order $\beta^0$
on the right hand side of the temperature derivative.
Further,
\begin{equation}
\omega_2
=
\frac{-\hbar^2}{4m} \nabla^2 U
- \frac{i\hbar}{2m}  {\bf p} \cdot \nabla U,
\end{equation}
and
\begin{eqnarray}
\omega_3  & = &
\frac{\hbar^2}{6m} \nabla U \cdot \nabla U
- \frac{\hbar^4}{24m^2} \nabla^2 \nabla^2 U
- \frac{i\hbar^3}{6m^2} {\bf p} \cdot \nabla \nabla^2 U
\nonumber \\ & & \mbox{ }
+ \frac{\hbar^2}{6m^2} {\bf p}{\bf p} : \nabla \nabla U.
\end{eqnarray}
The recursion relation is
\begin{eqnarray}
\lefteqn{
\omega_{n+1}
}  \\
& = &
\frac{- \hbar^2}{2(n+1)m}  (\nabla^2 U) \omega_{n-1}
- \frac{\hbar^2}{(n+1)m} \nabla U \cdot \nabla \omega_{n-1}
\nonumber \\ & & \mbox{ }
 + \frac{\hbar^2}{2(n+1)m} (\nabla U\cdot \nabla U) \omega_{n-2}
+ \frac{\hbar^2}{2(n+1)m} \nabla^2 \omega_{n}
\nonumber \\ & & \mbox{ }
+ \frac{i\hbar}{(n+1)m} {\bf p} \cdot  \nabla \omega_{n}
- \frac{i\hbar}{(n+1)m}  {\bf p} \cdot (\nabla U) \omega_{n-1} .\nonumber
\end{eqnarray}

\subsection{Series Expansion for $W$}

The partial differential equation (\ref{Eq:dW/dbeta})
gives series expansions for the effective potential
commutation function  $W$.\cite{STD2,Attard18b}
In powers of Planck's constant define
\begin{equation} \label{Eq:w=wnhn}
W \equiv \sum_{n=1}^\infty W_n \hbar^n ,
\end{equation}
with the classical limit being $W(\hbar=0)=0$.
(An expansion in powers of inverse temperature
is slightly simpler.)

The recursion relation for $n > 2$ is
\begin{eqnarray}
\frac{\partial W_n}{\partial \beta }
& = &
\frac{i}{m} {\bf p} \cdot \nabla W_{n-1}
 + \frac{1}{2m}
 \sum_{j=0}^{n-2}  \nabla W_{n-2-j}  \cdot \nabla W_j
\nonumber \\ && \mbox{ }
- \frac{\beta}{m} \nabla W_{n-2}  \cdot \nabla U
 + \frac{1}{2m} \nabla^2 W_{n-2} .
\end{eqnarray}

It is straightforward if somewhat tedious
to derive the first several coefficient functions explicitly.
One has
\begin{equation} \label{Eq:w1}
 W_1
=
\frac{-i\beta^2}{2m} {\bf p} \cdot \nabla  U ,
\end{equation}
\begin{eqnarray} \label{Eq:w2}
W_2
& = &
\frac{\beta^3}{6m^2}
{\bf p} {\bf p} : \nabla \nabla  U
 + \frac{1}{2m}
\left\{ \rule{0cm}{0.4cm}
\frac{ \beta^3}{3}  \nabla  U \cdot \nabla  U
-\frac{ \beta^2}{2}  \nabla^2 U
\rule{0cm}{0.4cm}\right\} ,
\nonumber \\ &&
\end{eqnarray}
\begin{eqnarray}
W_3
& = &
\frac{i\beta^4}{24m^3}
{\bf p} {\bf p} {\bf p} \vdots \nabla\nabla \nabla  U
+ \frac{5i\beta^4}{24m^2}  {\bf p}  (\nabla  U) : \nabla \nabla  U
\nonumber \\ && \mbox{ }
-\frac{i\beta^3}{6m^2} {\bf p} \cdot \nabla \nabla^2 U ,
\end{eqnarray}
and
\begin{eqnarray}
\lefteqn{
W_4
}  \\
& = &
\frac{- \beta^{5}}{5!m^4} ( {\bf p} \cdot \nabla )^4 U
- \frac{3\beta^5}{40m^3}   
(\nabla U ){\bf p} {\bf p} \vdots \nabla \nabla \nabla U
\nonumber \\ && \mbox{ }
-\frac{\beta^5}{15m^2}
 (\nabla U)  (\nabla U) :  \nabla\nabla  U
+ \frac{\beta^4}{16m^2}  \nabla U \cdot \nabla \nabla^2 U
\nonumber \\ && \mbox{ } \rule{0cm}{.8cm}
 + \frac{\beta^4}{16m^3}  
{\bf p} {\bf p} : \nabla \nabla \nabla^2 U
+ \frac{\beta^4}{48m^2}
\nabla^2 (\nabla  U \cdot \nabla  U)
 \nonumber \\ && \mbox{ }
- \frac{ \beta^3}{24m^2}  \nabla^2 \nabla^2 U
- \frac{\beta^5}{15m^3}  
( {\bf p} \cdot \nabla \nabla U ) \cdot ( {\bf p} \cdot \nabla \nabla U ) .
\nonumber 
\end{eqnarray}
(Here $W_4$ is taken from Ref.~[\onlinecite{Attard18b}],
which corrects Eq.~(7.112) of Ref.~[\onlinecite{STD2}].)

%
\section{Numerical Results} \label{Sec:Results}
\setcounter{equation}{0} \setcounter{subsubsection}{0}
\renewcommand{\theequation}{\Alph{section}.\arabic{equation}}
%

Since the original publication of the
many-body expansion proposed in this text,
after many trials and tribulations
I have finally succeeded in implementing the idea numerically.
I have tested it for the case of a one-dimensional
Lennard-Jones fluid in a simple harmonic oscillator potential.
Those results, which are not promising, are reported here.

In general integrating the temperature derivative
of the commutation function
from the high temperature limit can be unstable.
This is the method advocated in the text.
(The instability in the quadrature is not limited
to the many-body expansion.)
Although it works for the simple harmonic oscillator,
it does not, for example, give reliable results
for the Lennard-Jones pair potential.
Experience in a variety of cases
has shown that obtaining the energy eigenstates
and summing over them is by far the most efficient
and reliable way to obtain the commutation function.
For this reason an alternative method was used
to obtain the singlet and pair commutation functions.

The singlet combined commutation function,
given in the text by the  partial differential equation
for the temperature derivative,
Eq.~(\ref{Eq:tw1j}),
can be obtained for a one-particle system, $N=1$,
from the fundamental definition
\begin{eqnarray}
e^{\tilde w^{(1)}(q_1,p_1)}
& = &
\frac{1}{\langle q_1 |p_1 \rangle}
e^{- \beta \hat{\cal H}^{(1)}(q_1) }
\langle q_1 |p_1 \rangle
\nonumber \\ & = &
\frac{1}{\langle q_1 |p_1 \rangle}
\sum_\ell e^{-\beta E_\ell^{(1)} }
\langle q_1   | \ell \rangle \, \langle \ell | p_1 \rangle  .
\end{eqnarray}
Here the singlet Hamiltonian operator is
$\hat{\cal H}^{(1)}(q) = (-\hbar^2/2m) \nabla^2 + u^{(1)}(q)$,
and the singlet energy eigenfunctions satisfy
$ \hat{\cal H}^{(1)} | \ell \rangle = E_\ell^{(1)} | \ell \rangle$.
For the simple harmonic oscillator potential these are known
analytically.

For the case of two particles,
the total combined commutation function is
$\tilde W^{(2)}(q_1,p_1;q_2,p_2)
= \tilde w^{(1)}(q_1,p_1)
+ \tilde w^{(1)}(q_2,p_2)
+ \tilde w^{(2)}(q_1,p_1;q_2,p_2)$.
Hence the pair combined commutation function
given in the text by the  partial differential equation
for the temperature derivative,
Eq.~(\ref{Eq:tw2jk}),
can instead be obtained for a two-particle system,
$N=2$, from the fundamental definition
\begin{eqnarray}
\lefteqn{
e^{\tilde w^{(2)}(q_1,p_1;q_2,p_2)}
}  \\
& = &
e^{ \tilde W^{(2)}(q_1,p_1;q_2,p_2)
-\tilde w^{(1)}(q_1,p_1) - \tilde w^{(1)}(q_2,p_2) }
\nonumber \\ & = &
\frac{
e^{-\tilde w^{(1)}(q_1,p_1)} e^{-\tilde w^{(1)}(q_2,p_2)}
}{
\langle q_1,q_2  |p_1,p_2 \rangle
}
e^{- \beta \hat{\cal H}^{(2)}(q_1,q_2) }
\langle q_1,q_2 | p_1,p_2 \rangle
\nonumber \\ & = &
\frac{
e^{-\tilde w^{(1)}(q_1,p_1)} e^{-\tilde w^{(1)}(q_2,p_2)}
}{
\langle q_1,q_2  |p_1,p_2 \rangle
}
\sum_\ell \! e^{-\beta E_\ell^{(2)} } \!
\langle q_1,q_2   | \ell \rangle  \langle \ell | p_1,p_2 \rangle
.\nonumber
\end{eqnarray}
Here the pair Hamiltonian operator is
$\hat{\cal H}^{(2)}(q_1,q_2) = (-\hbar^2/2m) [\nabla_1^2 + \nabla_2^2 ]
+ u^{(1)}(q_1) + u^{(1)}(q_2) + u^{(2)}(q_1,q_2)$.
Again trial and much error has shown that it is most efficient
to obtain the pair combined commutation function
from the sum over energy states.
The latter were obtained  by standard minimization
and orthogonalization techniques.
The Fourier transform methods proposed in the text do not work
for the Lennard-Jones potential.

Two points: The system these were applied to consists of a singlet
simple harmonic oscillator potential
and a Lennard-Jones pair potential in one-dimension.
For the two-particle eigenfunctions in this system
the factorization into center of mass
and interaction coordinates is exact.
This means that one has to solve two one-dimensional systems
(one of which is the already solved simple harmonic oscillator case),
rather than a two-dimensional system.
Second, for the interaction coordinate,
the Lennard-Jones core repulsion makes the energy eigenfunction
vanish at $q=0$,
and the simple harmonic oscillator potential
makes it vanish at large $q$.
Hence one can choose a system size $q_\mathrm{max}$,
with $q \in [0,q_\mathrm{max}]$
and set $\phi_{\mathrm{int},\ell}(q_\mathrm{max})
= \phi_{\mathrm{int},\ell}'(q_\mathrm{max}) = 0$.
(This is done by setting the values beyond the boundary to zero
in the central difference formulae for the Laplacian.)
Then one can impose periodic boundary conditions,
and extend the eigenfunctions found on  $q \in [0,q_\mathrm{max}]$
to negative values, taking them to have the same energy
for the even and odd extensions.

\begin{figure}[t!]
\centerline{
\resizebox{8cm}{!}{ \includegraphics*{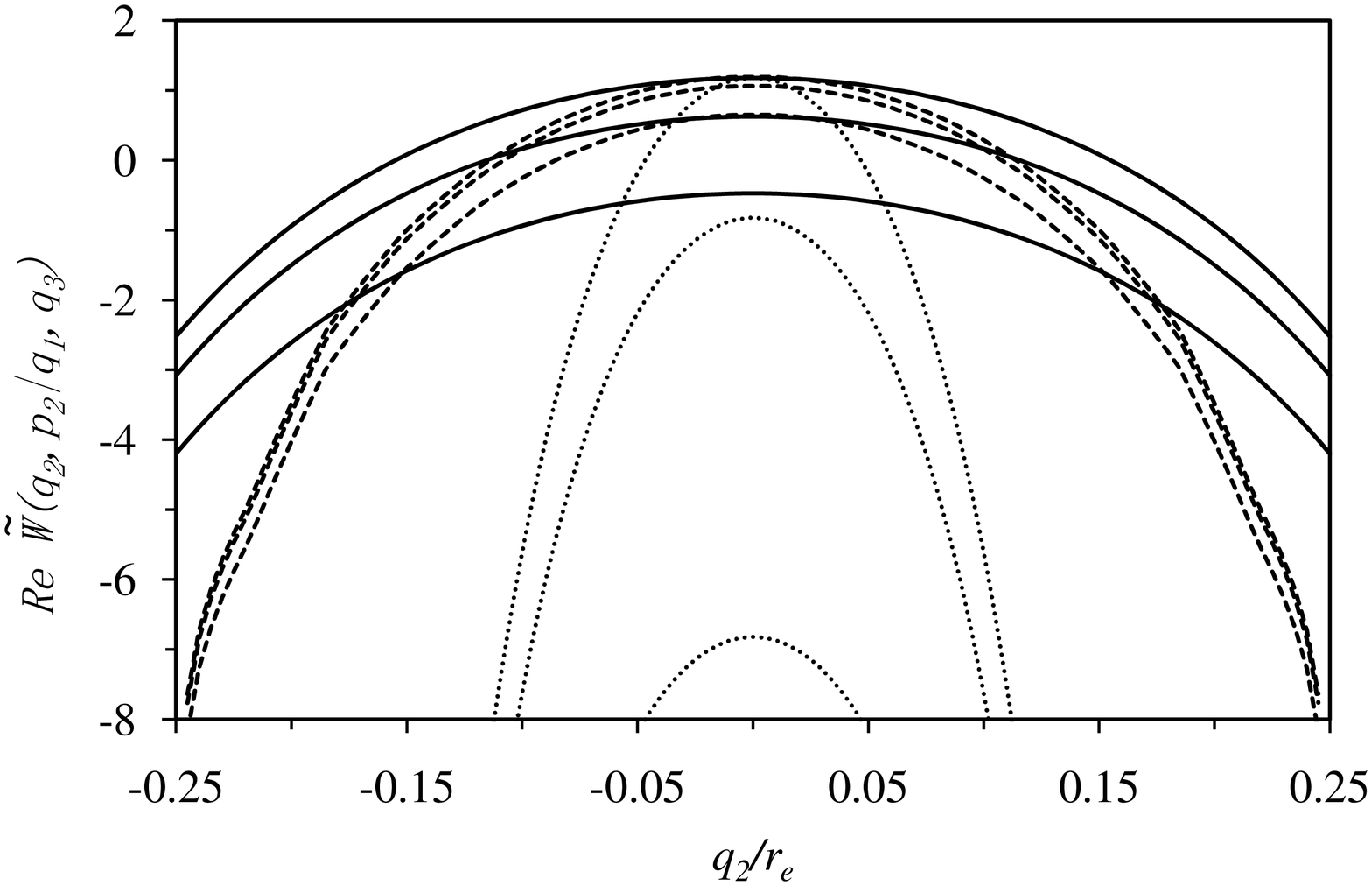} } }
\caption{\label{Fig:tildeW}
Real part of the combined total commutation function,
$\tilde W = W - \beta {\cal H}$,
for three particles, with $q_1 = - q_3 = -r_\mathrm{e}$
and $p_1 = p_3 = 0$,
at $\beta \hbar \omega = 0.5$
(Lennard-Jones pair interaction,
simple harmonic oscillator singlet potential,
one dimension).
The solid curves are the many-body expansion (pair),
the dashed curves are the local state expansion (triplet),\cite{Attard20}
and the dotted curves are the classical Maxwell-Boltzmann exponent,
$-\beta {\cal H}({\bf q},{\bf p})$.
From top to bottom in each series, the classical kinetic energy
for the middle particle is $\beta p_2^2/2m=$ 0, 2, and 8.
An arbitrary normalization constant has been added to each surface
so that they coincide at $\{q_2,p_2\}=\{0,0\}$.
}
\end{figure}

Figure~\ref{Fig:tildeW} shows the real part of the exponent
for the phase space weight for three particles,
$\mbox{Re } \tilde W^{(3)}(q_1,p_1;q_2,p_2;q_3,p_3)$,
with the two exterior particles fixed at $\pm r_\mathrm{e}$
with zero momentum.
Three approaches are compared:
the present many-body expansion (pair level),
a subsequently published local state expansion
(triplet level; one variable and two fixed particles),\cite{Attard20}
and the classical Maxwell-Boltzmann exponent.
(The imaginary part of the exponent,
which also contributes to the real part of the phase space weight,
is not shown.)
Overall, the data in Fig.~\ref{Fig:tildeW}
reveal that the two quantum approaches qualitatively agree
in that there is substantial cancelation of the classical core repulsion
between the Lennard-Jones particles.
This means that the quantum particles will approach each other
and interpenetrate much more deeply than their classical counterparts.
The pair many-body expansion
shows more cancelation than the triplet local state expansion.
The two approaches also agree in predicting
that higher values of momentum
are more accessible in the quantum case than is classically predicted.
However, for this, the real part of the exponent,
the many-body expansion shows greater variation with momentum
than does the local state expansion.
Since the Monte Carlo simulation results
for the average energy given by the local state expansion
appear to be in good agreement with benchmark results
for this system,\cite{Attard20}
the quantitative disagreement
between the present many-body expansion (pair level)
and the subsequent local state expansion (triplet level)\cite{Attard20}
does not bode well for the present approach.

\begin{figure}[t!]
\centerline{
\resizebox{8cm}{!}{ \includegraphics*{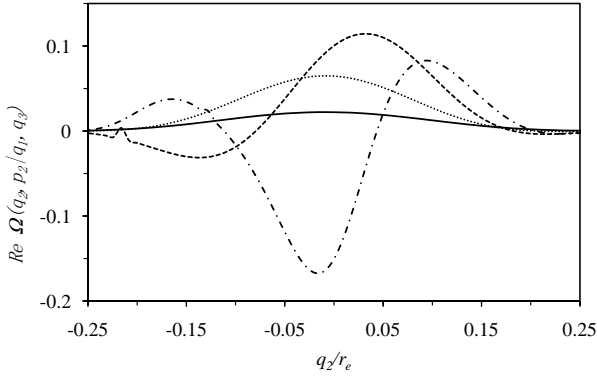} } }
\caption{\label{Fig:Omega}
The real part of the phase space weight
given by the many-body expansion (pair level)
for three particles, with $q_1 = -r_\mathrm{e}$,
$ p_1 r_\mathrm{e}/\hbar=8.38$  ($\beta p_1^2/2m = 5 $),
and $q_3 = r_\mathrm{e}$,
$ p_3 r_\mathrm{e}/\hbar=-11.85$  ($\beta p_3^2/2m = 10 $)
at $\beta \hbar \omega = 0.5$
(Lennard-Jones pair interaction,
simple harmonic oscillator singlet potential,
one dimension).
The solid curve is $ p_2 r_\mathrm{e}/\hbar=0$,
the dotted curve is $ p_2 r_\mathrm{e}/\hbar=5.24$,
the dashed curve is $ p_2 r_\mathrm{e}/\hbar=10.5$,
and
the dash-dotted curve is $ p_2 r_\mathrm{e}/\hbar=18.7$.
}
\end{figure}

Figure~\ref{Fig:Omega} shows the real part of the phase space weight,
$\mbox{Re } \Omega^{(3)}({\bf q},{\bf p}) =
\mbox{Re } \exp \tilde W^{(3)}({\bf q},{\bf p})$,
when the neighboring particles have non-zero momentum.
The results seem unphysical. In particular, the increase in magnitude
of the density with increase in momentum does not seem to be correct.
It appears to be the source
of an unrealistically high average kinetic energy
when the pair many-body expansion commutation function
is used in the Monte Carlo simulations (not shown).
The dramatic oscillations with position and with momentum,
and large negative values,
also seem unphysical, particularly considering
their relatively large magnitude.
Although the local state expansion also gives region of phase space
with negative weight,\cite{Attard20}
the magnitude of the density in such regions is quite small
relative to the maximum magnitude,
and so for that approach they either cancel or contribute negligibly
to any phase space average.

One can obtain a feeling for the limitations
of the pair terminated many-body expansion
by writing out the $N=3$ case  explicitly.
In this case Eq.~(\ref{Eq:DW.DW}),
in which the singlet terms are absent, yields
\begin{eqnarray}
\lefteqn{
\nabla \tilde W^{(3)} \cdot \nabla \tilde W^{(3)}
} \nonumber \\
& = &
\nabla_1 \tilde w^{(2)}_{12} \!\cdot\! \nabla_1 \tilde w^{(2)}_{12}
+ 2 \nabla_1 \tilde w^{(2)}_{12} \!\cdot\! \nabla_1 \tilde w^{(2)}_{13}
+ \nabla_1 \tilde w^{(2)}_{13} \!\cdot\! \nabla_1 \tilde w^{(2)}_{13}
\nonumber \\ & & \!
+ \nabla_2 \tilde w^{(2)}_{21} \!\cdot\! \nabla_2 \tilde w^{(2)}_{21}
+ 2 \nabla_2 \tilde w^{(2)}_{21} \!\cdot\! \nabla_2 \tilde w^{(2)}_{23}
+ \nabla_2 \tilde w^{(2)}_{23} \!\cdot\! \nabla_2 \tilde w^{(2)}_{23}
\nonumber \\ & & \!
+ \nabla_3 \tilde w^{(2)}_{31} \!\cdot\! \nabla_3 \tilde w^{(2)}_{31}
+ 2 \nabla_3 \tilde w^{(2)}_{31} \!\cdot\! \nabla_3 \tilde w^{(2)}_{32}
+ \nabla_3 \tilde w^{(2)}_{32} \!\cdot\! \nabla_3 \tilde w^{(2)}_{32}
\nonumber \\ & \approx &
\nabla_1 \tilde w^{(2)}_{12} \!\cdot\! \nabla_1 \tilde w^{(2)}_{12}
+ \nabla_2 \tilde w^{(2)}_{21} \!\cdot\! \nabla_2 \tilde w^{(2)}_{21}
+ \nabla_2 \tilde w^{(2)}_{23} \!\cdot\! \nabla_2 \tilde w^{(2)}_{23}
\nonumber \\ & & \mbox{ }
+ \nabla_3 \tilde w^{(2)}_{32} \!\cdot\! \nabla_3 \tilde w^{(2)}_{32}
+ 2 \nabla_2 \tilde w^{(2)}_{21} \!\cdot\! \nabla_2 \tilde w^{(2)}_{23} .
\end{eqnarray}
The second approximation retains only the nearest neighbor contributions.
One sees that the result in the text for the many-body expansion
terminated at the pair level, Eq.~(\ref{Eq:DW.DW2}),
neglects the final term here.
There is no reason to suppose that the neglected term
is smaller in magnitude than those that are retained.
(The commutation function is complex;
the retained terms are not in general positive.)

One can conclude from this that for more than two particles,
the many-body expansion of the total commutation function
terminated at the pair level
is inconsistent with its temperature derivative
based on the above approximation to the non-linear term.
It would mean, for example, that the commutation function
at one temperature, $W^{(N)}({\bf q},{\bf p};\beta)$,
formed using the series of singlet and pair commutation functions
obtained by summing over energy states at that temperature,
would not be equal to that obtained by
integrating the total commutation function from a higher temperature,
$W^{(N)}({\bf q},{\bf p};\beta_0)$,
using the approximation for the temperature derivative.
Alternatively,
the correct temperature derivative
implies the existence of three-body terms,
which contradicts the termination of the many-body expansion
at the pair level.
This inconsistency does not appear to be negligible,
and is no doubt responsible for the unphysical results
exhibited in the  figures above.

This result,
together with those in the figures
and the unpublished simulations,
suggest that the many-body expansion
advocated in the text is not viable at the pair level,
and perhaps not more generally.
It appears that  the local state expansion\cite{Attard20}
represents the path forward for quantum statistical mechanics
in classical phase space.


\begin{thebibliography}{99}
\bibitem{FN1}
In Ref.~[\onlinecite{Attard19}]
the symmetrization factor was shown correct for fermion statistics via
a proof by induction that the number of even and odd permutations
of $N$ objects were equal.
A much simpler proof is as follows:

Suppose that $\{\hat{\mathrm P}_\alpha\}_{\alpha=1}^{N!}$
is any set of permutation operators giving all the
 $N!$ permutations of $N$ objects.
Let $\hat{\mathrm P}_\alpha' = \hat{\mathrm P}_{jk}\hat{\mathrm P}_\alpha$,
where $\hat{\mathrm P}_{jk}$ transposes objects $j$ and $k$.
Clearly $\hat{\mathrm P}_\alpha'$ has opposite parity to
$\hat{\mathrm P}_\alpha$.
Since $\{\hat{\mathrm P}_\alpha'\}_{\alpha=1}^{N!}$
is also a set giving all the $N!$ permutations of $N$ objects,
it follows that the number of even permutations of $N$ objects
must equal the number of odd permutations.
\end{thebibliography}
\end{document}